\documentclass[preprint]{elsarticle}

\newtheorem{theorem}{Theorem}
\newtheorem{lemma}[theorem]{Lemma}

\newtheorem{definition}[theorem]{Definition}

\newproof{pf}{Proof}

\usepackage{amsmath}
\usepackage{color}
\usepackage{graphicx}
\usepackage{amssymb}

\journal{Finite Fields and Their Applications}

\begin{document}

\begin{frontmatter}
\title{New Families of $p$-ary Sequences of Period $\frac{p^n-1}{2}$ With Low Maximum Correlation Magnitude}

\author[cor1]{Wijik Lee}
\author{Ji-Youp Kim}
\author{Jong-Seon No}

\address[cor1]{Department of ECE, INMC, Seoul National University, Korea, tel number : 82-11-8568-7387, fax number : 82-2-880-8222, email : leewj422@ccl.snu.ac.kr}

\begin{abstract}
Let $p$ be an odd prime such that $p \equiv 3\;{\rm mod}\;4$ and $n$ be an odd integer. In this paper, two new families of $p$-ary sequences of period $N = \frac{p^n-1}{2}$ are constructed by two decimated $p$-ary m-sequences $m(2t)$ and $m(dt)$, where $d = 4$ and $d = (p^n + 1)/2=N+1$. The upper bound on the magnitude of correlation values of two sequences in the family is derived using Weil bound. Their upper bound is derived as $\frac{3}{\sqrt{2}} \sqrt{N+\frac{1}{2}}+\frac{1}{2}$ and the family size is $4N$, which is four times the period of the sequence.
\end{abstract}

\begin{keyword}
Character \sep correlation \sep m-sequences \sep $p$-ary sequences \sep Weil bound.
\end{keyword}

\end{frontmatter}

\section{Introduction}
Pseudorandom sequences with low correlation are widely used in wireless communications, that is, code division multiple access, spread spectrum, cryptography, and error correcting codes. Many papers of the sequence families with good correlation properties have been published. Kasami \cite{r1}, \cite{r2} proposed a binary sequence family with the optimal correlation property with respect to the Welch's lower bound. Further, there are lots of research results for the nonbinary sequence families. Liu and Komo \cite{r3} generalized the Kasami sequence family to $p$-ary case and Kumar and Moreno \cite{r4} constructed a $p$-ary sequence family with correlation magnitude upper bounded by $1+\sqrt{p^n}$ using bent function. Muller \cite{r5} also proposed two $p$-ary sequence families, whose correlation magnitude is upper bounded by $1+2\sqrt{p^n}$ and $1+\sqrt{p^n}$, respectively. Seo, Kim, No, and Shin \cite{r5.1} derived the cross-correlation distribution of $p$-ary sequences which have good correlation property. Choi, Lim, No, and Jung \cite{r5.2} also proposed a $p$-ary sequence family with correlation magnitude upper bound $\frac{p+1}{2}\sqrt{p^n}$ and family size $\sqrt{p^n}$.

Recently, $p$-ary sequence families with half period, that is, $N=\frac{p^n-1}{2}$ have been proposed. Generally half period sequences can have larger family size. Kim, Choi, and No \cite{r6} constructed $p$-ary sequence family of half period using Kloosterman sum. This sequence family has large family size of $4N$ and their correlation magnitude is upper bounded by $2\sqrt{N+\frac{1}{2}}$ for an odd prime $p \equiv 3 \,({\rm mod} \,\,4)$ and an odd integer $n$. And this result is further generalized by Kim, Chae, and Song \cite{r7}, that is, they generalized this sequence family to all odd prime $p$. Xia and Chen \cite{r8} constructed new sequence families having family size $4N$ and the correlation magnitude upper bounded by $\frac{1}{2}\left(p^e \sqrt{p^n+\frac{1}{p^{2e}}}\right)$.

Weil bound for exponential sums is often used to prove the upper bound on the magnitude of correlation values \cite{r9}. There are three types of Weil bounds. The first one is sum of multiplicative character. The second one is sum of additive character, and the last one is sum of multiplication of additive and multiplicative characters (hybrid type). Han and Yang \cite{r10} used multiplicative characters of Weil bound to derive the upper bound on the magnitude of correlation values. Wang and Gong \cite{r11} constructed polyphase sequence families whose correlation magnitude is derived from the Weil bound of exponential sums. They applied all three types of Weil bounds to the proof of the upper bounds.

In this paper, new $p$-ary sequence families with low correlation are constructed. For an odd prime $p\equiv 3 \,\,({\rm mod} \;\;4)$ and an odd integer $n$, two new $p$-ary sequence families of period $N = \frac{p^n-1}{2}$ having the correlation magnitude upper bounded by $\frac{3}{\sqrt{2}} \sqrt{N+\frac{1}{2}}+\frac{1}{2}$ are constructed. These sequence families can be obtained from shift and addition of two decimated $p$-ary m-sequences by 2 and $d$. One sequence family is obtained for $d=4$ and the other sequence family is constructed for $d = N+1$. The hybrid sum of Weil bound is used for the proof of the upper bound of correlation magnitude.

\section{Preliminaries}
This section introduces some basic definitions and concepts used in this paper. \\

\textit{A. Notations and Definitions}

1) Let $p$ be an odd prime such that $p \equiv 3\;{\rm mod}\;4$ and $n$ be an odd integer, where $q=p^n$. 

2) Let $\mathbb{F}_{q}$ be the finite field with $q$ elements and $\alpha$ be a primitive element of $\mathbb{F}_{q}$.

3) The trace function from $\mathbb{F}_{q}$ to $\mathbb{F}_{p}$ is defined as
\begin{equation*}
{\rm Tr}_{1}^{n}(x)=\sum_{i=0}^{n-1} x^{p^i}.
\end{equation*}

4) $\omega=e^{\frac{2\pi i}{p}}$ is a primitive complex $p$th root of unity, where $i=\sqrt{-1}$.

5) For some $\beta\in\mathbb{F}_{q}^*$, a $p$-ary m-sequence of period $q-1$ is defined as
\begin{equation*}
m(t) = {\rm Tr}_{1}^{n}(\beta\alpha^t).
\end{equation*}

6) Let $a(t)$ and $b(t)$ be $p$-ary sequences of period $N$. A cross-correlation between $a(t)$ and $b(t)$ is defined as
\begin{equation*}
C_{a,b}(\tau)=\sum_{t=0}^{N-1} \omega ^{a(t)-b(t+\tau)}.
\end{equation*}
If $a=b$, then the cross-correlation function becomes the autocorrelation function, denoted by $C_a(\tau)$. Let $S$ be a family of sequences of period $N$. Then the maximum magnitude of correlation values of the sequences in $S$ is defined as
\begin{equation*}
C_{\rm max}={\rm max}\left\{ \left| C_{a,b}(\tau) \right|: a,b\in S, 0\le \tau \le N-1, \tau \neq 0 \;\;{\rm if}\;\; a = b \right\}.
\end{equation*}

\textit{B. Characters and Weil Bound}

There are two types of characters, that is, additive character and multiplicative character as follows \cite{r12}.
\begin{definition}
(Additive Character): For $\beta \in \mathbb{F}_{q}$, an additive character of $\mathbb{F}_{q}$ is defined as\\
\begin{equation*}
\psi(x)=e^{\frac{2{\pi}i{\rm Tr}_{1}^{n}(\beta x)}{p}}, x\in\mathbb{F}_{q}
\end{equation*}
and $\psi _0$, $\psi(x)$ with $\beta = 0$, denotes the trivial additive character such that $\psi _0(x)=1$ for all $x\in\mathbb{F}_q$. 
\end{definition}

\begin{definition}
(Multiplicative Character): Let $g$ be a fixed primitive element of $\mathbb{F}_{q}$. For each $j=1,2,{\cdots},q-2,$ a multiplicative character of $\mathbb{F}_{q}$ is defined as
\begin{equation*}
\chi(g^k)=e^{\frac{2{\pi}ijk}{q-1}}
\end{equation*}
where $\chi (0)=0$ and $\chi _0$, $\chi(g^k)$ with $j=0$, denotes the trivial multiplicative character such that $\chi _0(x)=1$ for all $x\in \mathbb{F}_q^*$. 
\end{definition}

We consider the quadratic character $\eta$ in this paper, which is defined as
\begin{align*}
        \eta (y) =
        \begin{cases}
            ~~1, & \mbox{if}~y~\mbox{is nonzero square in}~\mathbb{F}_{q}\\
            -1, & \mbox{if}~y~\mbox{is nonzero nonsquare in}~\mathbb{F}_{q}\\
            ~~0, & \mbox{if}~y=0.
        \end{cases}
\end{align*}

\begin{lemma}
(Gaussian sum \cite{r12}): Let $\psi$ be an additive character of $\mathbb{F}_{q}$ and $\chi$ be a multiplicative character of $\mathbb{F}_{q}$. Then the Gaussian sum $G(\psi, \chi)$ is defined as
\begin{equation*}
G(\psi,\chi)=\sum_{x\in\mathbb{F}_{q}} \psi(x) \chi(x),
\end{equation*}
which satisfies
\begin{align*}
    G(\psi, \chi) =
    \begin{cases}
        p^n-1& \mbox{for}~\psi=\psi_0~\mbox{and}~ \chi=\chi_0\\
        0& \mbox{for}~\psi=\psi_0~\mbox{and}~  \chi \not= \chi_0\\
        -1& \mbox{for}~\psi \not= \psi_0~\mbox{and}~  \chi = \chi_0
    \end{cases}
\end{align*}
and for $\psi \neq \psi_0$ and $\chi \neq \chi_0$,
\begin{equation*}
\left| G(\psi, \chi) \right|=q^{1/2}.
\end{equation*}

\end{lemma}

The following Weil bounds are often used to prove the correlation property of the sequence.

\begin{theorem}
(Weil bound \cite{r9}): Let $\psi$ be a nontrivial additive character of $\mathbb{F}_{q}$ and $\chi$ be a nontrivial multiplicative character of $\mathbb{F}_{q}$ with order $M$ and $\chi(0)=0$. Let $f(x) \in \mathbb{F}_{q}[x]$ with degree $e$ and $g(x) \in \mathbb{F}_{q}$ with $s$ distinct roots in ${\overline{\mathbb{F}}_{q}}$, where $g(x)\neq c\cdot {h}^{M}(x)$ for some $c \in \mathbb{F}_{q}$ and $h(x) \in \mathbb{F}_{q}[x]$, and ${\overline{\mathbb{F}}_{q}}$ denotes the algebraic closure of $\mathbb{F}_q$. Then
\begin{equation*}
\left| \sum_{x\in\mathbb{F}_{q}} \chi(g(x))\psi(f(x)) \right| \le (e+s-1){\sqrt{q}}.
\end{equation*}

\end{theorem}

\begin{theorem}
(Additive type of Weil bound \cite{r12}): Let $f\in\mathbb{F}_q[x]$ be of degree $n\ge 1$ with ${\rm gcd}(n, q)=1$ and let $\psi$ be a nontrivial additive character of $\mathbb{F}_q$. Then
\begin{equation*}
\left|\sum_{x\in\mathbb{F}_{q}} \psi(f(x))\right| \le (n-1)\sqrt{q}.
\end{equation*}

\end{theorem}

\section{New Sequence Families and Their Correlation Bound}
In this section, we will propose two new $p$-ary sequence families of period $N=\frac{p^n-1}{2}$ and family size $4N$ and derive their correlation bound.
Let $m(t)$ be a $p$-ary m-sequence of period $q-1$. We consider the sequence $m(2t)$ and $m(dt)$, where $d=4$ and $N+1$. Since $q-1$ is even, the decimated sequence $m(2t)$ has the period $N$. Since ${\rm{gcd}}(q-1,d)=2$ for the both cases, the period of $m(dt)$ is also $N$. Then we define the new $p$-ary sequence family of period $N$ and family size $4N$ as

\begin{equation*}
S=\{m(2t+i)+m(d(t+l)+j)|0\le i,j\le 1, 0\le l\le N-1\}.
\end{equation*}

We will show that the magnitude of cross-correlation and nontrivial autocorrelation values of the sequences in the family $S$ is upper bounded by $\frac{\sqrt{3}}{2}\sqrt{N+\frac{1}{2}}+\frac{1}{2}$. For the proof of the upper bound, we use Theorems 4 and 5.

The correlation function between two sequences in $S$, $m(2t+i_1)+m(d(t+l_1)+j_1)$ and $m(2t+i_2)+m(d(t+l_2)+j_2)$, except for the trivial autocorrelation $(\tau =0, i_1 =i_2, j_1 =j_2, l_1 =l_2),$ is given as
\begin{align*}
C(\tau) &= \sum _{t = 0}^{N-1} \omega^{{\rm Tr}_{1}^{n}(\alpha ^{2t+i_1})+{\rm Tr}_{1}^{n}(\alpha ^{d(t+l_1)+j_1})-{\rm Tr}_{1}^{n}(\alpha ^{2(t+\tau)+i_2})-{\rm Tr}_{1}^{n}(\alpha ^{d(t+\tau+l_2)+j_2})}\\
&= \sum _{t = 0}^{N-1} \omega^{{\rm Tr}_{1}^{n}(\alpha ^{2t}(\alpha ^{i_1}-\alpha ^{2\tau +{i_2}})+\alpha ^{dt}(\alpha ^{dl_1 +j_1}-\alpha ^{d\tau +dl_2 +{j_2}}))}.
\end{align*}
Let $a = \alpha ^{i_1}-\alpha ^{2\tau +{i_2}}$ and $b = \alpha ^{dl_1 +j_1}-\alpha ^{d\tau +dl_2 +{j_2}}$. Then
\begin{equation*}
C(\tau)=\sum _{t=0}^{N-1} \omega ^{{\rm Tr}_{1}^{n}(a\alpha ^{2t}+b \alpha^{dt})}.
\end{equation*}

We will derive the upper bound of $C_{\rm max}$ for $d=4$ and $N+1$ in the following two theorems.

\begin{theorem}
: For $d=4$, we have
\begin{equation*}
C(\tau)=\sum _{t=0}^{N-1} \omega ^{{\rm Tr}_{1}^{n}(a\alpha ^{2t}+b \alpha^{4t})}.
\end{equation*}

Then, the maximum magnitude of $C(\tau)$ is given as
\begin{equation*}
C_{\rm max} \le \frac{3}{\sqrt{2}}\sqrt{N+\frac{1}{2}}+\frac{1}{2}.
\end{equation*}
\end{theorem}
\begin{pf}
Let $x = \alpha ^{2t}$ and QR be the set of quadratic residues of $\mathbb{F}_q$. Then we have
\begin{align}
C(\tau) &= \sum _{x\in\rm QR} \omega^{{\rm Tr}_{1}^{n}(ax + bx^2)}\nonumber \\
&=\frac{1}{2}\left(\sum _{x\in \mathbb{F}_{q}^{*}} \omega^{{\rm Tr}_{1}^{n}(ax + bx^2)}+\sum _{x\in \mathbb{F}_{q}^{*}} \eta(x) \omega^{{\rm Tr}_{1}^{n}(ax + bx^2)}\right).
\end{align}
Since the trivial autocorrelation case is excluded, it is easy to check that $a=b=0$ should not be considered because $i_1, i_2, j_1, j_2 \in \{0,1\}$.

(i) $b = 0$ and $a\neq 0$:\\
In this case, (1) can be rewritten as
\begin{equation}
\frac{1}{2}\left(\sum _{x\in \mathbb{F}_{q}^{*}} \omega^{{\rm Tr}_{1}^{n}(ax)}+\sum _{x\in \mathbb{F}_{q}^{*}} \eta(x) \omega^{{\rm Tr}_{1}^{n}(ax)}\right).
\end{equation}
The first term in (2) is given as
\begin{equation}
\sum _{x\in \mathbb{F}_{q}^{*}} \omega^{{\rm Tr}_{1}^{n}(ax)} = -1.
\end{equation}
Let $\chi = \eta, g(x) = x$, and $f(x) = ax$ in Theorem 4. Then the second term in (2) is computed as
\begin{equation}
\left|\sum _{x\in \mathbb{F}_{q}^{*}} \eta(x) \omega^{{\rm Tr}_{1}^{n}(ax)}\right| \le \sqrt{q}.
\end{equation}
From (3) and (4), (2) can be computed as
\begin{align}
\left|C(\tau)\right| = \frac{1}{2}\left|\left(\sum _{x\in \mathbb{F}_{q}^{*}} \omega^{{\rm Tr}_{1}^{n}(ax)}+\sum _{x\in \mathbb{F}_{q}^{*}} \eta(x) \omega^{{\rm Tr}_{1}^{n}(ax)}\right)\right| &\le \frac{\sqrt{q}+1}{2}\nonumber \\
&=\frac{\sqrt{2N+1}}{2}+\frac{1}{2}\nonumber \\
&=\frac{1}{\sqrt{2}}\sqrt{N+\frac{1}{2}}+\frac{1}{2}.
\end{align}
(ii) $b \neq 0$:\\
From Theorem 5 with $f(x)=ax+bx^2$, the first term in (1) can be derived as
\begin{equation}
\left|\sum _{x\in \mathbb{F}_{q}^{*}} \omega^{{\rm Tr}_{1}^{n}(ax + bx^2)}\right| \le \sqrt{q}+1.
\end{equation}
Let $\chi = \eta, g(x) = x$, and $f(x) = ax+bx^2$ in Theorem 4. Then, the second term in (1) is computed as
\begin{equation}
\left|\sum _{x\in \mathbb{F}_{q}^{*}} \eta(x) \omega^{{\rm Tr}_{1}^{n}(ax+bx^2)}\right| \le 2\sqrt{q}.
\end{equation}
From (6) and (7), we have
\begin{align}
\frac{1}{2}\left|\left(\sum _{x\in \mathbb{F}_{q}^{*}} \omega^{{\rm Tr}_{1}^{n}(ax+bx^2)}+\sum _{x\in \mathbb{F}_{q}^{*}} \eta(x) \omega^{{\rm Tr}_{1}^{n}(ax+bx^2)}\right)\right| &\le \frac{3}{2}\sqrt{q}+\frac{1}{2} \nonumber \\
&=\frac{3}{\sqrt{2}}\sqrt{\frac{q}{2}}+\frac{1}{2}=\frac{3}{\sqrt{2}}\sqrt{N+\frac{1}{2}}+\frac{1}{2}.
\end{align}
From (5) and (8), we prove the theorem.
\end{pf}
\begin{theorem}
: Let $d=N+1$. $C(\tau)$ can be rewritten as
\begin{equation}
C(\tau)=\sum _{t=0}^{N-1} \omega ^{{\rm Tr}_{1}^{n}(a\alpha ^{2t}+b \alpha^{(N+1)t})}.
\end{equation}
Then, the maximum magnitude of $C(\tau)$ can also be derived as
\begin{equation*}
C_{\rm max} \le \frac{3}{\sqrt{2}}\sqrt{N+\frac{1}{2}}+\frac{1}{2}.
\end{equation*}
\end{theorem}
\begin{pf}
Let $x = \alpha ^{t}$. It is easy to check that $-1$ is a nonsquare in $\mathbb{F}_{p^n}$ for an odd integer $n$ and an odd prime $p\equiv 3 \;({\rm mod} \;4)$. Let $x = y^2$ for a square $x$ and $x=-y^2$ for a nonsquare $x$.\\
Since $N+1$ is even, we have the same form of
\begin{equation*}
        {\rm Tr}_{1}^{n}(ax^{2}+bx^{\frac{N+1}{2}}) = {\rm Tr}_{1}^{n}(ay^{4}+by^{2})
\end{equation*}
for both $x=y^2$ and $-y^2$.
Then (9) can be rewritten as
\begin{align}
C(\tau)&=\sum _{y\in\mathbb{F}_q^*} \omega ^{{\rm Tr}_{1}^{n}(ay^{4}+by^{2})}\nonumber\\
&=\frac{1}{2}\left(\sum _{y\in \mathbb{F}_{q}^{*}} \omega^{{\rm Tr}_{1}^{n}(ay^2 + by)}+\sum _{y\in \mathbb{F}_{q}^{*}} \eta(y) \omega^{{\rm Tr}_{1}^{n}(ay^2 + by)}\right).
\end{align}
Since (10) is the same as (1) by swapping $a$ and $b$, the proof is the same as Theorem 6.
Thus the proof is done.
\end{pf}

\begin{theorem}
: The family size of $S$ is $4N$.
\end{theorem}
\begin{pf}
If there are two cyclically equivalent sequences in $S$, then their cross-correlation value is equal to $N$. From Theorems 6 and 7, the magnitude of the cross-correlation values of arbitrary two sequences are upper bounded by $\frac{3}{\sqrt{2}}\sqrt{N+\frac{1}{2}}+\frac{1}{2}$ and thus the sequences in $S$ are cyclically inequivalent.
\end{pf}

Even though the maximum magnitude of correlation values of the proposed sequence families is upper bounded, the number of distinct correlation values increases as $N$ becomes large. Table I shows the number of distinct correlation values and the normalized maximum magnitude of $C_{\rm max}$ by $\sqrt{N}$ for some $p$ and $n$. In case of $p=3$ and odd $n$, the number of distinct correlation values is less than 6 and the correlation distribution is researched in \cite{r8}.

\begin{table}[t]
\begin{center}
\centering
\caption {
Simulation results of ${C_{\rm max}}$ and number of correlation values for some $p$ and $n$.}
\begin{tabular}{|c|c|c|c|c|}\hline
$p$ & $n$ & $N$ & $\frac{C_{\rm max}}{\sqrt{N}}$ & Number of distinct values \\ \hline
  & 3 & 13 & 2.1650 & 5 \\ \cline{2-5}
3 & 5 & 121 & 2.1259 & 6 \\ \cline{2-5}
  & 7 & 1093 & 2.1219 & 6 \\ \cline{2-5}
  & 9 & 9841 & 2.1214 & 6 \\ \hline 
7 & 3 & 171 & 2.0304 & 94 \\ \cline{2-5}
  & 5 & 8403 & 2.0951 & 852 \\ \hline
11 & 3 & 665 & 2.0003 & 450 \\ \hline
\end{tabular}
\end{center}
\end{table}

\section{Conclusion}
In this paper, for an odd integer $n$ and an odd prime $p$ such that $p \equiv 3\;{\rm mod}\;4$, two new families of $p$-ary sequences with low maximum correlation magnitude are constructed where the period of sequences is $N=\frac{p^n-1}{2}$ and the family size $4N$. The sequences in the family are obtained using shift and additions of the decimated $p$-ary m-sequences $m(2t)$ and $m(dt)$, where $d = 4$ and $N+1$. The upper bound for the magnitude of cross-correlation and nontrivial autocorrelation values of the sequences in the family $S$ can be evaluated as $\frac{3}{\sqrt{2}}\sqrt{N+\frac{1}{2}}+\frac{1}{2}$ using the Weil bound and the family size is four times the period of sequences, $4N$.

\end{document}